\documentstyle[12pt,epsf]{article}

\begin{document}

\def\bb{b\bar{b}}
\def\bu{B^+}
\def\bd{B^0_d} 
\def\bs{B^0_s}
\def\lb{\Lambda_b}
\def\bmix{B^0 \mbox{--} \overline{B^0}}
\def\bdmix{B_d^0 \mbox{--} \overline{B_d^0}}
\def\bsmix{B_s^0 \mbox{--} \overline{B_s^0}}
\def\bsg{b\to s\,g}
\def\dmd{\Delta m_d}
\def\dms{\Delta m_s}
\def\kstar{K^{\ast 0}}
\def\kstarbar{\overline{K}^{\ast 0}}
\def\sinsqth{\sin^2\theta_W^{eff}}
\def\Zbb{Z^0 \rightarrow b\,{\overline b}}
\def\Zcc{Z^0 \rightarrow c\,{\overline c}}
\def\Zff{Z^0 \rightarrow f\,{\overline f}}
\def\Zqq{Z^0 \rightarrow q\,{\overline q}}
\def\Zuds{Z^0 \rightarrow u\,{\overline u},d\,{\overline d},s\,{\overline s}}


\pagestyle{empty}

\renewcommand{\thefootnote}{\fnsymbol{footnote}}
                                                  

\begin{flushright}
{\small
SLAC--PUB--8598\\
August 2000\\}
\end{flushright}
                
\begin{center}
{\bf\Large Time Dependent $B_s^0 - \overline{B_s^0}$ Oscillations
Using Exclusively Reconstructed $D_s^+$ Decays
at SLD\footnote{Work supported in part by the
Department of Energy contract  DE--AC03--76SF00515.}}

\bigskip

The SLD Collaboration$^{**}$
\smallskip

Stanford Linear Accelerator Center, \\
Stanford University, Stanford, CA 94309\\
\medskip

\vspace{1.5cm}

{\bf\large
Abstract }
\end{center}
\noindent
We set a preliminary 95\% C.L. exclusion  on the oscillation frequency of
$B_s^0 - \overline{B_s^0}$ mixing using
a sample of 400,000 hadronic $Z^0$ decays collected
by the SLD experiment at the SLC during the 1996-98 run.
In this analysis, $B_s^0$ mesons are partially reconstructed by combining
a fully reconstructed $D_s$ with other $B_s^0$ decay tracks.  The 
$D_s$ decays are reconstructed via the $\phi \pi$ and $K^* K$ channels.
The $b$-hadron flavor at production is determined by exploiting the
large forward-backward asymmetry of
polarized $Z^0 \rightarrow b \overline{b}$ decays
as well as information
from the hemisphere opposite to the reconstructed $B$ decay.
The flavor of the $B_s^0$ at the decay vertex is determined by the
charge of the $D_s$.
A total of 361 candidates passed the final event selection cuts.  
This analysis excludes the following values of the
$B_s^0 - \overline{B_s^0}$ mixing oscillation frequency:
$\Delta m_s < 1.5$ ps$^{-1}$,
$2.6 < \Delta m_s < 4.9$ ps$^{-1}$, and 
$10.8 < \Delta m_s < 13.5$ ps$^{-1}$ at the 95\% confidence level.

\vspace{2cm}

\begin{center}

{\sl Paper Contributed to the XXXth International Conference on
     High Energy Physics (ICHEP 2000), Osaka, Japan, 27 July - 2 August, 2000.}

\end{center}
\vfill

\normalsize

\pagebreak
\pagestyle{plain}

\pagebreak

\section{Introduction}
  The Standard Model allows $B^0 \leftrightarrow \overline{B^0}$ oscillations
to occur via second order weak interactions.  The frequency of
oscillation is determined by the mass differences, $\Delta m$, 
between the mass eigenstates in the $B^0$ system.
The mass difference in the $B^0_s$ system ($\Delta m_s$) and 
in the $B^0_d$ system ($\Delta m_d$)
are proportional to the Cabibbo-Kobayashi-Maskawa (CKM) matrix elements  
$\left| V_{ts} \right|$ and $\left| V_{td} \right|$, respectively.
A measurement of $\Delta m_d$ can in principle be used to extract the
CKM matrix element $\left| V_{td} \right|$.
However, the extraction of $\left| V_{td} \right|$ from $\Delta m_d$ 
is complicated by a large theoretical uncertainty on the hadronic
matrix elements.  The complication can be circumvented by taking the ratio
of $\Delta m_s$ and $\Delta m_d$.  In the ratio, the theoretical uncertainties
partially cancel, giving
\begin{equation}
{\frac{\Delta m_s}{\Delta m_d}}={\frac{m_{B_s^0}}{m_{B_d^0}}}\xi^2
\left| \frac{V_{ts}}{V_{td}} \right|^2,
\end{equation}
where $\xi^2$ is estimated from Lattice QCD calculation to be 
$ (1.11^{+0.06}_{-0.04})^2$~\cite{bstocchi,bhash} and $m_{B_s^0}$ 
($m_{B_d^0}$) is the mass of the $B_s^0$ ($B_d^0$) meson.
Therefore, a direct measurement of $\Delta m_s$, combined with the
current measurement of $\Delta m_d$, can be translated to 
a value of $\left| V_{td} \right|$ with a 5$\%$ precision.  This has
motivated many experiments to search for $B_s^0$
oscillations.  As of now, all analyses have failed to observe
significant signal and only lower limits on $\Delta m_s$ have been
set.

A $B_s^0$ mixing analysis requires several key ingredients: a
$B_s^0$ enriched event sample, knowledge of the $B_s^0$ meson flavor at
the production and decay vertex, and the $B_s^0$ proper decay time.
In this paper, an analysis to search for $B_s^0$ oscillations
using fully reconstructed $D_s$ is presented.  We begin in section
2 with a brief description of the experimental apparatus.  
In section 3, the details of the event selection and $Ds$ reconstruction
are discussed.  The two $Ds$ decay channels used in the analysis are: 
\begin{equation}
B_s^0 \rightarrow D_s^- + X, \ \ \ \ D_s^- \rightarrow \phi \ \pi^-,
\ \ \ \ \ \ \ \ \phi \rightarrow K^+ K^- ; {\ \ \ \ \ \ \ \ \ } 
\end{equation}
\begin{equation}
B_s^0 \rightarrow D_s^- + X, \ \ \ \ D_s^- \rightarrow K^{*0} K^-, 
\ \ \ \ K^{*0} \rightarrow K^+ \pi^- . \ \ \ \ \ \ \ 
\end{equation}
The next two sections describe the method for reconstructing the
$B_s^0$ proper decay time and determining the flavor of the $B_s^0$ at the 
production and the decay vertices.  Finally, the fitting procedure and the 
results are discussed in the remaining sections.  The result presented 
in this paper is based on the data collected during the 1996-1998 runs which
consist of 400,000 hadronic $Z^0$ decays with an average longitudinal
electron beam polarization of about 73$\%$.

\section{The Detector}

The Stanford Large Detector (SLD) is a general purpose particle detector
designed to study the decay of $Z^0$ boson produced at the Stanford Linear 
Collider (SLC).  A detailed description of the detector can be found
here~\cite{bsld,bvxd3}.  In this section, the main features of
the detector components that are important to the analysis are summarized.

At the heart of the SLD is a CCD pixel vertex detector (VXD3) with over 
300 million pixels.  It consists of 3 concentric layers at radii of
2.8, 3.8 and 4.8 cm.  The polar angle coverage of the vertex detector  
extends from $cos(\theta)$ of -0.85 to 0.85.
The single hit resolution is roughly $4 \mu{m}$ in 
each coordinate and the measured track impact parameter resolutions are: 
\begin{displaymath}
\sigma_{r \phi}=7.8 \oplus \frac{33}{(p sin^{3/2}\theta)} \mu{m} \ , 
\ \ \ \
\sigma_{r z}=9.7 \oplus \frac{33}{(p sin^{3/2}\theta)} \mu{m} \ , 
\end{displaymath}
where z axis points along the beampipe and the track momentum ($p$)
 is measured in GeV/c.
The VXD3 is surrounded by the Central Drift
Chamber (CDC) that extends from the inner radius of 20cm to the outer
radius of 100cm.  Reconstructed tracks from the CDC are linked with 
tracks from the vertex detector to improve the momentum resolution.  
The momentum resolution for the combined tracks (CDC+VXD3) is measured to 
be
\begin{displaymath}
{\sigma_{p_{\perp}}}/{p_{\perp}}=
{0.0095}\oplus 0.0026{p_{\perp}} \ ,
\end{displaymath}  
where $p_{\perp}(GeV/c)$ is the momentum transverse to the beampipe.  
The next layer of detector is the main particle identification 
system at SLD, the Cherenkov Ring Imaging Detector (CRID).  The CRID comprises
two radiator systems ($C_6F_{14}$ liquid and $C_5F_{12}$ gas) for 
$K / \ \pi$ separation over a large momentum range (0.35-35.0 GeV/c).
The barrel CRID provide particle identification over the central
70\% of the solid angle. 
Immediately outside the CRID is the Liquid Argon Calorimeter (LAC), which
is used primarily for triggering, energy reconstruction, and electron 
identification.  The
calorimeter towers are made of planes of lead plates separated by 
non-conducting spacers and immersed in liquid argon.  The energy resolutions
for electromagnetic showers is measured to be 
$\sigma_{e.m.} / {E}={15\%} / \sqrt{E(GeV)}$, 
and for the hadronic showers is estimated to be
$\sigma_{had}/{E}={60\%}/ \sqrt{E(GeV)} $.

The above four central sub-detectors
are inside a solenoidal magnet that provides a uniform axial magnetic field 
of 0.6 Tesla.  
The outermost detector is the Warm Iron Calorimeter (WIC).  The WIC consists
of 14 layers of 5cm thick iron plates instrumented with streamer tubes 
between layers for muon identification.  

\section{Event Selection and $D_s$ Reconstruction}

\subsection{$b \ \overline{b}$ Event Selection}
\begin{figure}[p]
  \centering
  \epsfxsize11cm
  \leavevmode
  \epsfbox{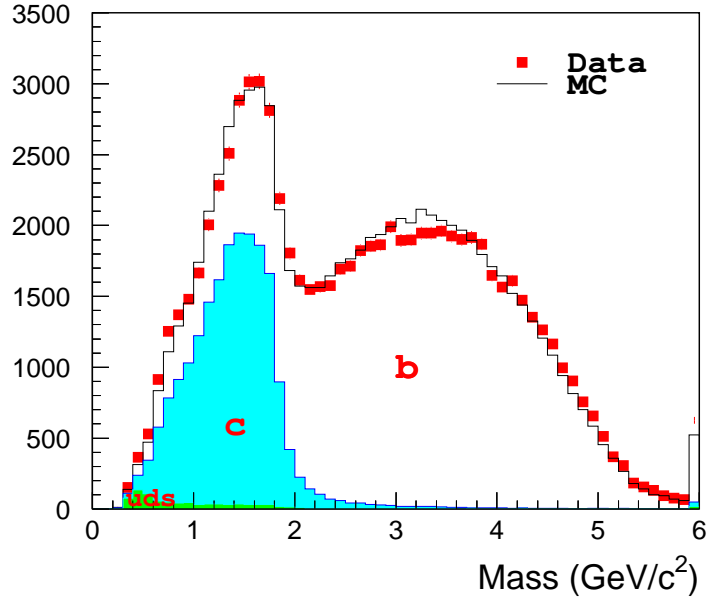}
  \caption{\it \label{fig:b3mass}
  \baselineskip=12pt
  $p_t$ corrected vertex mass distribution for data (red square) and
MC (solid line).  White histogram is $b\overline{b}$ events,
blue histogram is $c\overline{c}$ events, and green histogram is uds 
events~\cite{bsudong}. }
  \baselineskip=18pt
\end{figure}

Events used in the analysis are first required to pass the hadronic
event selection.   An event is selected as a hadronic candidate if it 
satisfies the following conditions: has at least 7 good CDC tracks
(each track is required to have momentum transverse to the beampipe
greater than 200MeV) that pass within 5 cm of the interaction point (IP),
the thrust axis is within $\left| cos(\theta_{thrust}) \right| < 0.85$,
and the total charged track energy is greater than 18 GeV.  The hadronic
event selection removes essentially all di-lepton events 
($e^+e^-\rightarrow \it{l^+}\it{l^-}$) and other
non-hadronic backgrounds.  To further enhance $Z^0 \rightarrow b \overline{b}$
events in the sample, hadronic candidates are required to have at least 
one topologically reconstructed secondary vertex~\cite{bzvtop} with a vertex
mass greater than 2 GeV that is corrected for the 
missing transverse momentum to partially account for the neutral 
particles.  A neural net is used in the vertex
mass reconstruction to enhance the separation between $b\overline{b}$ 
and other $q\overline{q}$ events in the sample~\cite{btom}.  
Figure~\ref{fig:b3mass} shows the $p_t$ corrected
vertex mass distribution for data and MC.  A minimum vertex mass cut at 2 GeV 
yields an event sample with b hadron purity of about 98$\%$ and single
hemisphere $b$ tagging efficiency of about 56$\%$.

\subsection{$D_s^-$ $\rightarrow$ $\phi \pi^- , K^{*0} K^-$ Reconstruction}
\begin{figure}[p]
  \centering
  \epsfxsize16cm
  \leavevmode
  \epsfbox{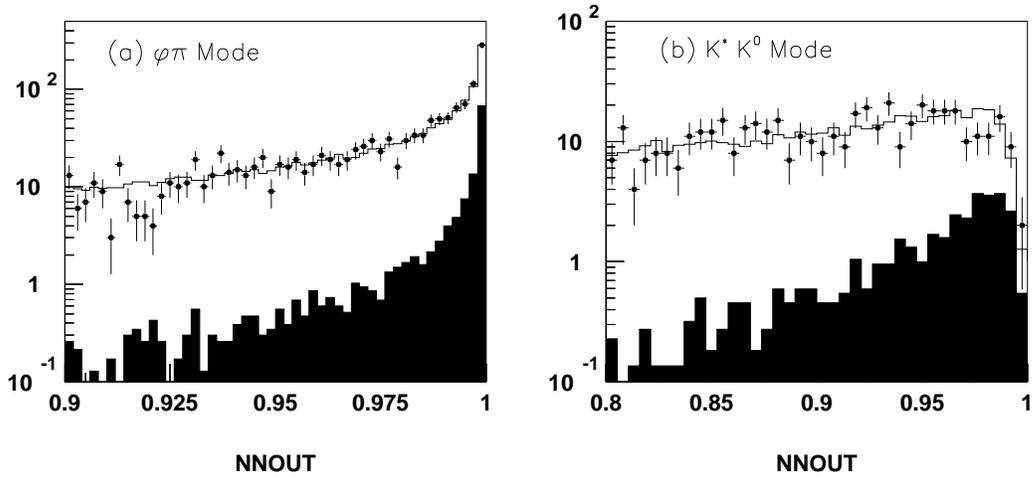}
  \caption{\it \label{fig:nnout}
  \baselineskip=4pt
  Neural Net outputs for (a) $\phi \pi$ and (b) $K^*K^0$ modes.  Dots are
  data, open histograms are Monte Carlo and  solid black histograms 
  represent MC simulation of true Ds events.}
  \baselineskip=15pt
\end{figure}

$D_s^-$ candidates are reconstructed by first pairing oppositely
charged tracks to form a $\phi$ ($K^{*0}$) candidate for the $\phi \pi^-$  
($K^{*0}K^-$) mode.  A third track is then attached to form a $D_s$
candidate.  Charged tracks used in the combination are required to
have at least 23 CDC hits (out of possible 80 hits), at least 2 vertex 
detector hits, and a combined
$\chi^2$/d.o.f. for the  CDC+VXD3 fit of less than 8 to ensure good 
reconstruction. 
To maximize the discrimination between true $D_s$ and combinatorial 
background events, kinematic information for the $D_s$ candidate is fed 
into a neural net.  The neural net inputs for
the $\phi \pi$ ($K^{*0} K$) mode include: $K^+K^-$ ($K^+\pi^-$) invariant 
mass, fitted vertex probability 
of the $D_s$, total momentum of the $D_s$, helicity angle $\theta^*$ 
(angle between the $\phi$ ($K^{*0}$) and the $D_s$ flight directions
in the $D_s$ rest frame), helicity angle $\lambda^*$ (angle between
the charged daughter of the $D_s$ and the $K^+$ from $\phi$ ($K^{*0}$) decay 
in the rest frame of the neutral meson), and 
particle ID information for the three tracks.  
The complete list of neural net inputs is shown in Table~\ref{tbl:dsnn}.
\begin{table}[hbt]
\begin{center}
\begin{tabular}{c||c} \hline
$\phi \pi $ \ Mode  & $K^{*0} K $ \ Mode \\  \hline \hline
$D_s$ vertex $\chi^2$ prob & $D_s$ vertex $\chi^2$ prob \\ \hline
$P_{ptot}(D_s)$ & $P_{ptot}(D_s)$ \\ \hline
$KK$ opening angle & $K\pi$ (from $K^{*0}$) opening angle \\ \hline
$D_s$ normalized decay length & $D_s$ normalized decay length \\ \hline
helicity angle $\lambda^*$  & helicity angle $\lambda^*$ \\ \hline
helicity angle $\theta^*$  & helicity angle $\theta^*$ \\ \hline
particle ID of three tracks & particle ID of three tracks \\ \hline
 & Average Normalized 3-D impact \\ 
 & parameter of $KK\pi$ tracks \\ \hline 
\end{tabular}
\caption {$D_s$ Neural net inputs - left column is for $\phi \pi$
and right column is for $K^{*0} K$}
\label{tbl:dsnn}
\end{center}
\end{table}
The neural net is trained on Monte Carlo events generated using
JETSET 7.4 \cite{bjetset} with full detector simulation based on 
GEANT 3.21 \cite{bgeant}.
The neural net outputs are shown in Figure~\ref{fig:nnout}.

\begin{figure}[p]
  \centering
  \epsfxsize11cm
  \leavevmode
  \epsfbox{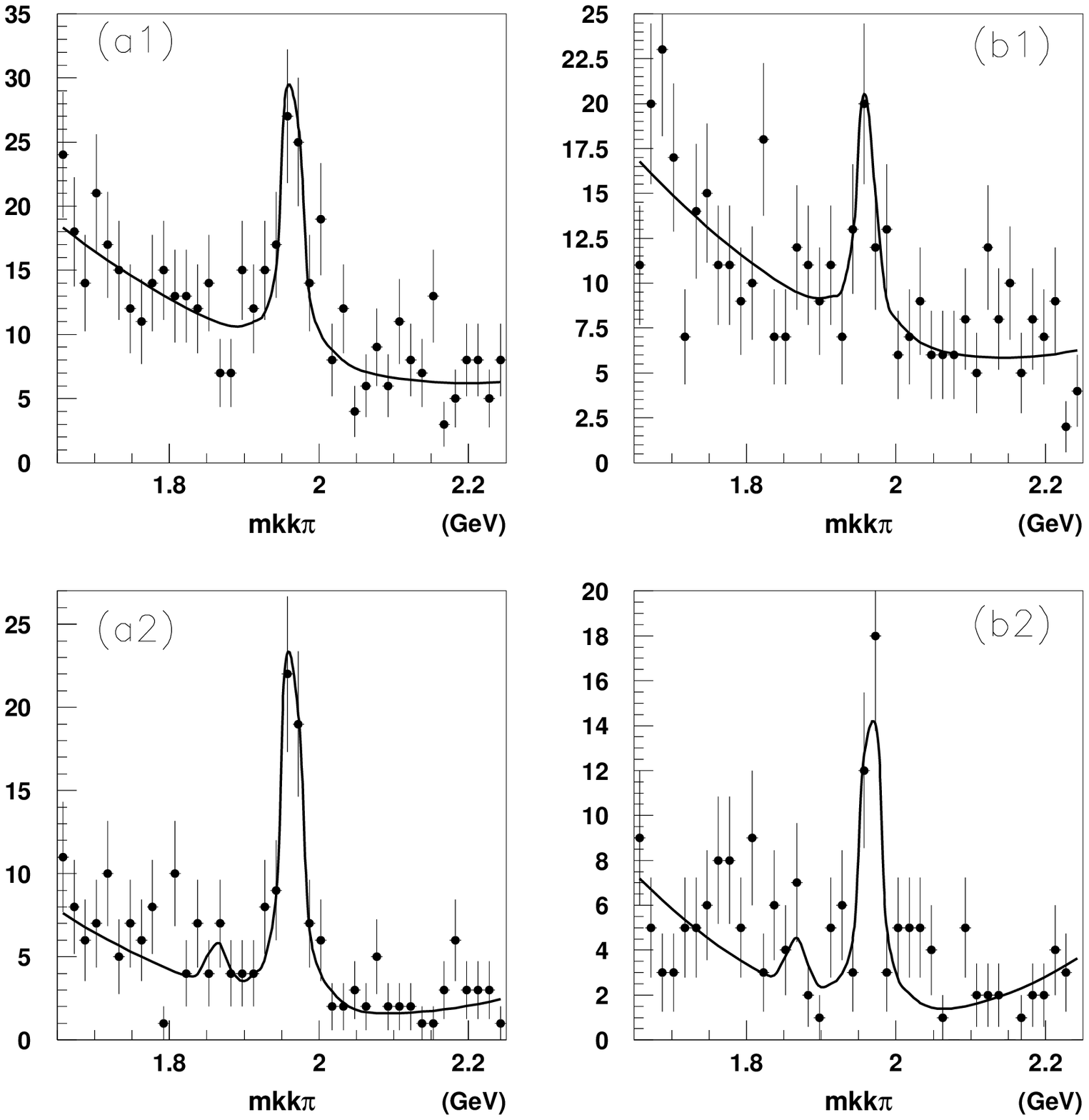}
  \caption{\it \label{fig:dsmass_phipi}
  \baselineskip=4pt
  Distribution of ${KK\pi}$ invariant mass for the $\phi \pi$ mode.  
  Plots in the left column 
  (a1,a2) are for the neutral candidates and plots in the right column 
  (b1,b2) are for the charged candidates. The $\phi \pi$ sample is
  further subdivided into events with loose kaon ID (a1,b1) and 
  events with hard kaon ID (a2,b2).}
  \baselineskip=15pt
\end{figure}

\begin{figure}[p]
  \centering
  \epsfxsize8cm
  \leavevmode
  \epsfbox{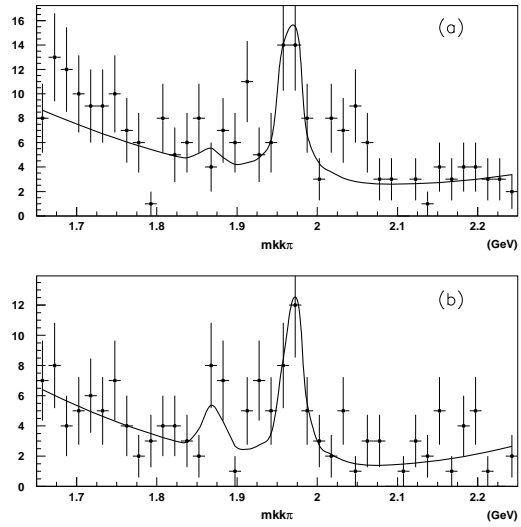}
  \caption{\it \label{fig:dsmass_kstark}
  \baselineskip=4pt
  Distribution of ${KK\pi}$ invariant mass for the $K^* K^0$ mode. 
 (a) for events with
  1 kaon ID and (b) for events with 2 kaons ID.}
  \baselineskip=15pt
\end{figure}

The optimal neural net cut that maximizes the sensitivity of the
analysis to $B_s^0$ mixing is determined separately for each of the
two $D_s$ decay modes.
The minimum cut for the $\phi \pi$ mode is determined to be at 0.9 and
for the $K^{*0} K$ mode, is at 0.8. 
Furthermore, for the $K^{*0} K$ mode, the kaon from the $K^{*0}$ decay is
required to be identified by the CRID in order to suppress combinatorial
and other non-$D_s$ backgrounds.  The $KK \pi$ invariant mass spectra
are shown in Figures~\ref{fig:dsmass_phipi} and~\ref{fig:dsmass_kstark}.  
The $D_s$ mass peak is fitted separately
for events with and without definite kaon ID and for Q=0 and Q=$\pm$1
events, where Q is defined as the total charge of all tracks associated
with the B decay.
The details of the B vertex reconstruction
and the $m_{KK\pi}$ mass fits are described in later sections.
The estimated number of $D_s$ candidates
and combinatorial fractions are given in Table~\ref{tbl:dsmass1} 
and~\ref{tbl:dsmass2}.
\begin{table}[hbt]
\begin{center}
\begin{tabular}{|c||c|c||c|c|} \hline
 & \multicolumn{2}{c||}{Without Definite Kaon ID } & 
\multicolumn{2}{|c|}{With Definite Kaon ID} \\ 
&\ \ \ Q=0 \ \ \ & Q=$\pm$1 &\ \ \ Q=0 \ \ \ & Q=$\pm$1 \\ \hline
\# of hadronic candidates &101  & 57  &54  & 40 \\
\# of semileptonic candidates &8  & 6  &12  & 2 \\
Average combinatorial fraction& 42.3$\pm$2.1$\%$ &53.8$\pm$2.7$\%$
 & 21.7$\pm$2.2$\%$  & 14.1$\pm$1.4$\%$  \\ \hline
\end{tabular}
\caption {$\phi \pi$ mode-- number of $D_s$ candidates within $\pm$ 40MeV
of the nominal mass and the estimated average combinatorial fractions.}
\label{tbl:dsmass1}
\end{center}
\end{table}
\begin{table}[hbt]
\begin{center}
\begin{tabular}{|c||c||c|} \hline
 &{\ \ 1 Kaon ID \ \ } & {2 Kaons ID} \\ 
 &\ \ \ Q=0 \ \ \ &\ \ \ Q=0 \ \ \ \\ \hline
\# of hadronic candidates & 40 & 30  \\
\# of semileptonic candidates & 6 &5   \\
Combinatorial fraction        & 32.9$\pm$3.3$\%$ &22.3$\pm$2.2$\%$  \\ \hline
\end{tabular}
\caption {$K^{*0} K$ mode-- number of $D_s$ candidates within $\pm$ 40MeV
of the nominal mass and the estimated average combinatorial fractions.}
\label{tbl:dsmass2}
\end{center}
\end{table}

\section{B Proper Time and Vertex Charge Reconstruction}
The $B_s^0$ oscillation frequency is expected to be large in the 
Standard Model and 
therefore a time dependent study of the phenomenon requires precision  
reconstruction of the $B_s^0$ proper decay time.  The proper decay time is 
calculated from the reconstructed decay length ($\it{l}$) and the 
momentum of the $B_s^0$:
\begin{equation}
\tau_{B_s}=\frac{\it{l}}{\beta\gamma c},
\end{equation} 
where $\beta\gamma$, often referred to as ``boost'', is the ratio between 
the momentum of the $B_s^0$ and its mass.  In the following sections,
the algorithms for reconstructing the decay length and the boost are
described.
\subsection{Decay Length Reconstruction}
The decay length of the $B_s^0$ is defined as the distance between
the IP and the $B_s^0$ decay vertex.  The IP position is found by vertexing
all tracks in the event to a common vertex.  Taking advantage of 
the stability of the SLC beamspot, the IP position in the x-y plane (transverse
to the beampipe) is averaged over 30 hadronic events.  The position in 
z (beampipe direction) is calculated on an event-by-event basis.
The estimated IP resolution for $\Zbb$ events is about 4$\mu m$ in x-y and 
20$\mu m$ in z.

The $B_s^0$ decay vertex is found by vertexing the $D_s$ track
with other $B$ decay tracks in the hemisphere.  This is accomplished
in two steps.  The first step involves the selection of the seed
vertex (preliminary estimate of the $B_s^0$ decay vertex).  To find the 
seed, the $D_s$ track is individually vertexed with each quality track 
(excluding $D_s$ daughters)  
in the same hemisphere, and
the vertex that is farthest from the IP and upstream (or consistent
with being upstream within 5$\sigma$) of the $D_s$ and has a vertex 
fit $\chi^2$ of  
less than 5 is chosen as the seed.  Step two involves the separation of
tracks into secondary decay and fragmentation tracks.  
The discriminating variable used is the L/D parameter, where
L is defined as the distance from the IP to the point  
of the closest approach of the candidate track to the seed vertex 
axis(line joining IP to the seed), and D is the distance from the 
IP to the seed.  A track is chosen as a secondary B decay
track if it satisfies the following two conditions:
1.) track L/D is greater than 0.5 and 2.) forms a 
good vertex with the $D_s$ track (fit $\chi^2 \le$ 5).  The latter condition
is imposed to reject spurious and B daughter tracks (from
double charm decays) that do
not point back to the B vertex.
 Finally, the selected tracks are then vertexed together 
with the $D_s$ 
to obtain the best estimate of the B decay position. 
The resulting B decay length resolution is highly dependent on the
decay topology.  
The list of decay length resolutions estimated from Monte Carlo for 
the various decay categories is shown in Table~\ref{tbl:sigl}.
\begin{table}[hbt]
\begin{center}
\begin{tabular}{c|c c|c c} \hline
{ } & \multicolumn{2}{c}{Q=0} & 
\multicolumn{2}{|c}{Q=$\pm$1} \\
Decay Category  & Core $\sigma_L$($\mu m$) & Tail $\sigma_L$($\mu m$) &
Core $\sigma_L$($\mu m$) & Tail $\sigma_L$($\mu m$) \\  \hline \hline
$B_s^0 \rightarrow D_s^- X$ (right-sign)& 47 & 144 & 51 &184 \\ \hline
$B_s^0 \rightarrow D_s^+ X$ (wrong-sign)& 89 & 292 & 89 &292 \\ \hline
$B_d^0 \rightarrow D_s^\pm X$ & 70 & 271 & 85 & 412 \\ \hline
$B_u \rightarrow D_s^\pm X$ & 99 & 435 & 72 & 236 \\ \hline
$B Baryon \rightarrow D_s^\pm X$ & 84 & 221 & 84 & 221 \\ \hline
\end{tabular}
\caption {Decay length resolutions for various decay topologies.  
Resolutions are parameterized separately for neutral and charged
B events except for wrong-sign $B_s^0$ and B Baryon events.  
The resolution is parameterized by the sum of two 
gaussians with core fraction fixed to 60$\%$.
\label{tbl:sigl}
}
\end{center}
\end{table}

\subsection{Boost Reconstruction}
To obtain the relativistic boost, the energy
of the B meson has to be reconstructed.
The total energy of the $B$ meson is the sum of the energy
of the charged and neutral daughter particles:
\begin{equation}
E^B=E^B_{charged}+E^B_{neutral}.
\end{equation}
The charged energy is determined by summing all the charged
tracks associated with the B decay assuming pion mass (except
for the two kaons from the $D_s$ decay).  To estimate the neutral 
energy contribution, five different techniques are used.  The first
four techniques are calorimetery based and use various
constraints (beam energy, jet energy, $B_s^0$ mass and calorimeter
information) to estimate
the neutral energy of the B meson~\cite{bmoore}.
The fifth technique is based only on the kinematics of the decay
(B vertex axis, charged track momentum and $B_s^0$ mass constraint) to 
estimate the neutral energy~\cite{bdong}. 
The results from the five algorithms are then averaged, taking correlations
into account, to obtain the total B energy.  
The relative boost resolutions 
$\sigma (\frac{\beta\gamma^{rec}-\beta\gamma^{true}}{\beta\gamma^{true}})$ 
are shown in Table~\ref{tbl:sigb}.
\begin{table}[hbt]
\begin{center}
\begin{tabular}{c|c c|c c} \hline
{ } & \multicolumn{2}{c}{Q=0} & 
\multicolumn{2}{|c}{Q=$\pm$1} \\
Decay Category  & Core $\sigma_{\beta\gamma}$($\%$) & 
Tail $\sigma_{\beta\gamma}$($\%$) &
Core $\sigma_{\beta\gamma}$($\%$) & 
Tail $\sigma_{\beta\gamma}$($\%$) \\  \hline \hline
$B_s^0 \rightarrow D_s^- X$ (right-sign)& 7.9 & 19.1 & 10.1 & 26.5 \\ \hline
$B_s^0 \rightarrow D_s^+ X$ (wrong-sign)& 8.6 & 19.7 & 11.1 & 28.3 \\ \hline
$B_d^0 \rightarrow D_s^\pm X$           & 9.5 & 19.6 &  9.1 & 24.7 \\ \hline
$B_u \rightarrow D_s^\pm X$             & 10.4& 30.0 &  8.4 & 21.1 \\ \hline
$B Baryon \rightarrow D_s^\pm X$        & 10.0& 25.4 & 11.1 & 39.6 \\ \hline
\end{tabular}
\caption {Relative boost resolutions for various decay topologies.  
The resolution is parameterized by the sum of two 
gaussians with core fraction fixed to 60$\%$.
\label{tbl:sigb}
}
\end{center}
\end{table}

\subsection {Charge Reconstruction}
\begin{figure}[htb]
    \centering
    \epsfxsize8cm
    \leavevmode
    \epsfbox{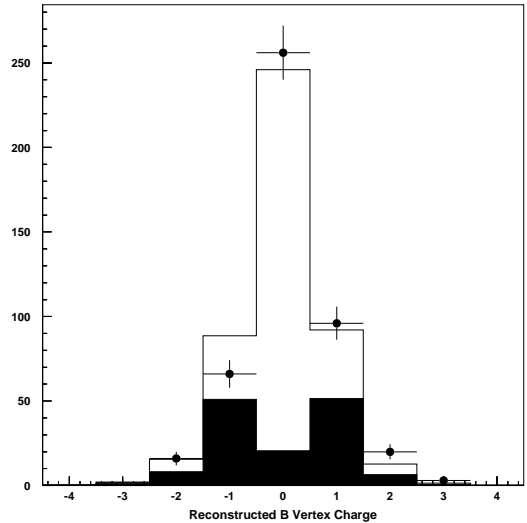}
    \caption{\it \label{fig:bvtxchg}
    \baselineskip=12pt
    Reconstructed B vertex charge distribution for data (dots),
    M.C. neutral B events (open histogram), and M.C. charged 
    B events (solid histogram). The distribution is for the $\phi \pi$
    and the $K^{*0}K$ modes combined.}
  \end{figure}

Nominally, the charge of the $b$ hadron is the sum of the charge of the
quality tracks
associated with the B decay.  To improve charge purity, tracks with
only VXD3 hits (VX-alone vectors) that are not used in vertex
and momentum reconstruction are also included in the B charge
determination~\cite{btom}.  For the $\phi \pi$ mode, both the
neutral B candidates (Q=0) and the charged candidates (Q=$\pm$1) are used.  
The $B_s^0$ purity in the charged sample is considerably lower 
than in the neutral sample and therefore has reduced weight in the fit.
For the $K^{*0} K$ mode, only the
neutral candidates are included in the final event sample.
The reconstructed vertex charge
distributions for data and Monte Carlo are shown in Figure~\ref{fig:bvtxchg}.

\section{Flavor Tagging}
To decide whether mixing has occured for a given event, the flavor 
of the $B_s^0$ has to be determined at 
the production and the decay point.
In this section, the methods used to tag the initial
and final state of the $B_s^0$ are described.

\subsection {Final State Flavor Tag}
A significant fraction of the $B_s^0$ decays result in a $D_s$ in
the final state. This knowledge is used not only to enhance 
the $B_s^0$ purity in the data sample but also as a means to determine 
the flavor of
the $B_s^0$ at the decay vertex.  For the final state tag,
we assume the $D_s$ to come from the $b \rightarrow c$ transition which
implies that a  $\overline{B_s^0}$ would decay into a $D_s^+$ and 
a $B_s^0$ would decay into a $D_s^-$.
However, roughly 10$\%$ of the time the virtual W produces a $D_s$ 
with an opposite sign.  This process is the dominant source of final
state mistag.  The issue of final state mistag will be further addressed
in a later section of this paper.


\subsection {Initial State Flavor Tag}
Several techniques are used to determine the initial 
state of the $B_s^0$.  The most powerful method, unique to the SLD, 
is the polarization tag.  In a 
polarized $Z^0 \rightarrow b\overline{b}$ decay, the 
outgoing quark is produced preferentially along the direction 
opposite to the spin of the $Z^0$ boson.  Therefore by 
knowing the helicity of the electron beam and the 
direction of the jet, the flavor of the primary quark 
in the jet can be inferred.  The analyzing power of 
the polarization tag is highly dependent on the polar angle of
the jet w.r.t. the beam axis and the electron beam 
polarization.  The probability that 
the jet is a b quark jet is
\begin{equation}
P(b)=\frac{1+A_{FB}}{2},
\end{equation}
where $A_{FB}$ is the polarized forward-backward 
asymmetry, defined as,
\begin{equation}
A_{FB}=2A_b \frac{A_e - P_e}{1-A_e P_e}
\frac{cos\theta_T}{1+cos^2\theta_T},
\label{eqn:afb}
\end{equation}
with $A_b$ and $A_e$ assumed to have the Standard Model
values of 0.935 and 0.150, respectively.  The polar
angle $\theta_T$ in equation~(\ref{eqn:afb}) is defined 
as the angle between the thrust axis, which points in
the event hemisphere, and the electron beam direction.
$P_e$ is the electron beam polarization ($P_e >$ 0 for right 
handed and $<$ 0 for left handed electron beam).

In addition to the polarization tag, the momentum
weighted jet charge technique is also used to 
determine the initial state of the $B_s^0$.
The quantity $Q_{jet}$ is defined as
the sum of the charge of the tracks in the opposite
hemisphere weighted by the longitudinal momentum
of the track w.r.t. the thrust axis, namely 
\begin{equation}
Q_{jet}=\sum_{i}^{trks}{\left| \vec{p}_i \cdot \vec{T} \right|^\kappa},
\end{equation}
where $\vec{T}$ is the thrust axis and $\kappa$ (equal to 0.5) 
is a parameter determined from Monte Carlo that maximizes the jet 
charge separation between $q$ and $\overline{q}$ jets.
The probability of tagging the b quark hemisphere, given
the opposite hemisphere jet charge $Q_{jet}$, is
\begin{equation}
P(b)=\frac{1}{1+e^{\alpha Q_{jet}}},
\end{equation}
where the coefficient $\alpha$ is determined from Monte Carlo
to be -0.27.

To further enhance the initial state tag purity, additional
sources of information from the opposite hemisphere are used.  
This includes: vertex charge, charge of lepton, 
charge of kaon and dipole charge~\cite{bdipo}.
All the available tags in a given event are combined to
obtain the overall initial state tag probability. For
this analysis, the average initial state correct tag
probability is about 75$\%$.  

\section{$B_s^0$ Oscillation Studies}
The final step in a mixing analysis is to fit for the
oscillation frequency.  This entails a search for periodic 
oscillations in the proper decay time distribution of events tagged
either as mixed or unmixed. We expect the proper time distribution, in the
limit that the lifetime difference between the two mass 
eigenstates is small, to have the following time dependence for 
the mixed and unmixed events:
\begin{equation}
P_{mixed}(\tau)=\frac{e^{-\tau/\tau_{B_s}}}{2\tau}\left(
1-cos(\Delta m_s\tau) \right),
\end{equation}
\begin{equation}
P_{unmixed}(\tau)=\frac{e^{-\tau/\tau_{B_s}}}{2\tau}\left(
1+cos(\Delta m_s\tau) \right).
\end{equation}
Where $\tau_{B_s}$ is the $B_s^0$ lifetime and $\Delta m_s$ is
the mass difference, as introduced in the earlier section.
The fitting function used in the analysis also
includes the effect of detector resolution, reconstruction 
efficiency, mistag, and background events.  These details
are addressed in the next few sections.

\subsection{Amplitude Fit}
The amplitude fit \cite{bmoser} is a fitting method that is used in this
analysis.
The method, in essence, transforms the traditional likelihood fit into
a ``Fourier-like'' analysis.
In an amplitude
fit, the likelihood function is modified by introducing a
term A, the amplitude, in front of the cosine terms
($\cos(\Delta m_s \tau) \rightarrow {\bf{A}}\cos(\Delta m_s \tau)$).
Instead of fitting for $\Delta m_s$ directly, $\Delta m_s$ is fixed
to a particular value and the parameter {\bf A} is fitted for. The fit
for {\bf A} is repeated for a range of $\Delta m_s$ values to produce
the amplitude plot.  The amplitude is expected to be consistent with zero for 
values of $\Delta m_s$ far from $\Delta m_s^{true}$ and 
to reach unity at the true mass difference value.  
If the oscillation frequency is large and no signal
is observed, the range of $\Delta m_s$ for which ${\bf A}+1.645\sigma_A
\le 1$ can be excluded at the 95$\%$ confidence level.  The 95$\%$ C.L.
sensitivity is defined as the value of $\Delta m_s$ at which 
$1.645\sigma_A=1$.
 
\subsection{Likelihood Function}
To perform the amplitude fit, we first need to construct
a likelihood function that describes the proper time distribution
of the events in the data sample.
The events in the final sample can be divided into seven main sources, each
with its own proper time distribution function, $F_x$.  
The seven physics sources are:
\begin{itemize}
   \item $F_{B_s}^{r.s.} = B_s^0 \rightarrow D_s^- X $ right sign decays + c.c.
   \item $F_{B_s}^{w.s.} = B_s^0 \rightarrow D_s^+ X $ wrong sign decays + c.c.
   \item $F_{B_d} = B_d^0 \rightarrow D_s^{\pm} X $ + c.c.
   \item $F_{B^{\pm}} = B^{\pm} \rightarrow D_s^{\pm} X $ + c.c.
   \item $F_{BB}$ = B Baryon $\rightarrow D_s^{\pm} X $ + c.c.
   \item $F_{cc}$ = primary charm quark $\rightarrow D_s^{+} X $ + c.c.
   \item $F_{comb}$ = combinatorial events.
\end{itemize}
The proper time distribution of the event sample is the sum of the 
seven physics functions with the contribution from each source
weighted by its fraction in the sample and the appropriate normalization
constant.  The resulting normalized probability distribution function 
for the mixed events is
\begin{eqnarray}
P_{mixed}(\tau_{rec})& = &  {\it f_{D_s}(m_{KK\pi})} \left(
\frac{{\it{f_{B_s}^{r.s.}}}}{N_1}F_{B_s}^{r.s.}+
\frac{{\it{f_{B_s}^{w.s.}}}}{N_2}F_{B_s}^{w.s.}+
\frac{{\it{f_{B_d}}}}{N_3}F_{B_d}+
\frac{{\it{f_{B^{\pm}}}}}{N_4}F_{B^{\pm}}+ \right. \nonumber \\
& & \left. \frac{{\it{f_{BB}}}}{N_5}F_{BB}+
\frac{{\it{f_{cc}}}}{N_6}F_{cc} \right) + 
\left[ 1-{\it f_{D_s}(m_{KK\pi})} \right] \cdot F_{comb} \ , \label{eqn:pmix}
\end{eqnarray}
where $\tau_{rec}$ is the reconstructed proper time, $\it{f_{D_s}}$ is the 
fraction of $D_s$ in the sample,  
$\it{f_{x}}$ is 
the fraction of category x in the $D_s$ signal peak, and $N_i$ is the 
normalization
constant for category {\it i}.  
The physics functions for the mixed events have the following form:
\begin{eqnarray}
\label{eqn:fbsrs}
F_{B_s}^{r.s.}(\tau_{rec}) & = & \int_{0}^{\infty}
{\frac{e^{-\tau/\tau_{B_s}}}{2\tau_{B_s}}\left[(1-\eta_i)
\left(1-A\cdot cos(\Delta m_s\tau) \right)+
\eta_i \left(1+A\cdot cos(\Delta m_s\tau) \right) \right]} \cdot \nonumber \\
& & \epsilon_1 (\tau) \cdot G_1(\tau, \tau_{rec}) \ d\tau . \\
F_{B_s}^{w.s.}(\tau_{rec}) & = & \int_{0}^{\infty}
{\frac{e^{-\tau/\tau_{B_s}}}{2\tau_{B_s}}\left[(1-\eta_i)
\left(1+A\cdot cos(\Delta m_s\tau) \right)+
\eta_i \left(1-A\cdot cos(\Delta m_s\tau) \right) \right]} \cdot \nonumber \\
& & \epsilon_2 (\tau) \cdot G_2(\tau, \tau_{rec}) \ d\tau . \\
F_{B_d}(\tau_{rec}) & = & \int_{0}^{\infty}
{\frac{e^{-\tau/\tau_{B_d}}}{2\tau_{B_d}}\left[(1-\eta_{B_d})
\left(1-cos(\Delta m_d\tau) \right)+
\eta_{B_d} \left(1+cos(\Delta m_d\tau) \right) \right]} \cdot \nonumber \\
& & \epsilon_3 (\tau) \cdot G_3(\tau, \tau_{rec}) \ d\tau . \\
F_{B^{\pm}}(\tau_{rec}) & = & \int_{0}^{\infty}
{\frac{e^{-\tau/\tau_{B^{\pm}}}}{2\tau_{B^{\pm}}}\cdot
\eta_{B^\pm}\cdot \epsilon_4 (\tau) \cdot G_4(\tau, \tau_{rec})} \ d\tau . \\
\label{eqn:fbb}
F_{BB}(\tau_{rec}) & = & \int_{0}^{\infty}
{\frac{e^{-\tau/\tau_{B_{BB}}}}{2\tau_{B_{BB}}}\cdot
\eta_{BB}\cdot \epsilon_5 (\tau) \cdot G_5(\tau, \tau_{rec})} \ d\tau . \\
F_{cc}(\tau_{rec})&=&5.47e^{-\tau_{rec}/0.2102}*erf(18.35\tau_{rec}) \ \ \ 
(parameterized \ from \ M.C.) . \\
F_{comb}(\tau_{rec})&=& (distribution \ taken \ from \ data \ sidebands).
\end{eqnarray}
Where, $\tau$ is the 
true proper time, $\eta_i$ is the initial state mistag probability, 
$\eta_{B_{d,\pm,B}}$ are the overall
mistag probabilities for $B_{d,\pm,B}$ events, and the last 
two functions: $\epsilon_{\it i}(\tau)$ and
$G_{\it i}$($\tau,\tau_{rec}$) are the vertex efficiency and the
resolution functions for category ${\it i}$, respectively.
The proper time distribution for the combinatorial events is taken
directly from the data using events in the sidebands.  Detailed M.C. 
studies have shown that the combinatorial distribution is well modelled
by the sidebands and that using the sideband distribution does not
introduce a bias in the amplitude fit. 
The corresponding probability distribution for the unmixed events 
is obtained by replacing $\eta$ with ($1-\eta$) in the physics
functions.  Finally, the unbinned likelihood function is defined as the 
product of the probabilitiy of the mixed and unmixed events
\begin{equation}
{\it{L}}= \prod_i^{mixed} P_{mxied}(\tau_{rec}) 
\prod_j^{unmixed} P_{unmxied}(\tau_{rec}).
\end{equation}

\subsubsection{Data Compositions and Event Mistag Rate}
The $D_s$ fraction ($\it{f_{D_s}}$) in equation~(\ref{eqn:pmix}) is
the fraction of true $D_s$ events in the sample.
Instead of using
the average value from the mass plot, the $D_s$ fraction is determined 
on an event-by-event basis using the reconstructed $D_s$ mass of the 
candidate event.
Events close to the nominal $D_s$ mass are more likely to be
true $D_s$ events than combinatorial events, therefore have more
significance in the analysis.  The $D_s$ signal and background 
parameterizations used in the $\it{f_{D_s}}$ calculation are 
taken directly from the $m_{KK\pi}$ mass plot.
The fitting function for the $D_s$ mass peak is the sum of two gaussians
with the same mean and 60$\%$ core fraction.   The widths used in the 
individual fits are fixed to the widths from the combined fit.  
The background is parameterized by a second order polynominal (combinatorial
background) and a gaussian function$(D^+$ mass peak).  For the $D^+$ gaussian,
only the amplitude is allowed to float; the mean is fixed to 
99.2MeV \cite{bpdg} below the fitted $D_s$ peak and the width is taken 
from the Monte Carlo distribution.  
The six dominant sources of $D_s$ production are outlined in the
beginning of this section.  The relative fractions of the sources 
are determined from the Monte Carlo and are estimated separately
for the charged and neutral samples as well as for hadronic and
semileptonic decays.  The semileptonic decays are defined as events 
where a B decay track (not from the $D_s$ decay) is identified 
as either a muon or an electron.  The relative fractions used in the fit 
are listed in Table~\ref{tbl:fracs1} and~\ref{tbl:fracs2}.
\begin{table}[hbt]
\begin{center}
\begin{tabular}{|c||c|c||c|c|} \hline
 & \multicolumn{2}{c||}{\ \ Hadronic Decays \ \ } & 
\multicolumn{2}{|c|}{Semileptonic Decays} \\ 
&\ \ \ Q=0 \ \ \ & Q=$\pm$1 &\ \ \ Q=0 \ \ \ & Q=$\pm$1 \\ \hline
${\it f_{B_s}^{r.s.}}$ &0.556  &0.283  &0.822  &0.452  \\
${\it f_{B_s}^{w.s.}}$ &0.066  &0.034  &0.036  &0.020  \\
${\it f_{B_d}}$        &0.260  &0.160  &0.122  &0.129  \\
${\it f_{B^{\pm}}}$    &0.046  &0.452  &0.013  &0.387  \\
${\it f_{BB}}$         &0.053  &0.052  &0.007  &0.011  \\
${\it f_{cc}}$         &0.019  &0.019  &0.     &0. \\ \hline
\end{tabular}
\caption {Fractions for $\phi \pi$ mode.}
\label{tbl:fracs1}
\end{center}
\end{table}
\begin{table}[hbt]
\begin{center}
\begin{tabular}{|c||c||c|} \hline
 &{\ \ Hadronic Decays \ \ } & {Semileptonic Decays} \\ 
 &\ \ \ Q=0 \ \ \ &\ \ \ Q=0 \ \ \ \\ \hline
${\it f_{B_s}^{r.s.}}$ &0.558  &0.766    \\
${\it f_{B_s}^{w.s.}}$ &0.055  &0.075    \\
${\it f_{B_d}}$        &0.252  &0.117    \\
${\it f_{B^{\pm}}}$    &0.048  &0.014    \\
${\it f_{BB}}$         &0.051  &0.028    \\
${\it f_{cc}}$         &0.036  &0.    \\ \hline
\end{tabular}
\caption {Fractions for $K^{*0} K$ mode.}
\label{tbl:fracs2}
\end{center}
\end{table}

The $B_s^0$ contribution to the likelihood function is divided
into the right-sign and wrong-sign decay terms, and therefore,  
by construction, the final state mistag rate is 0 and 1, respectively,
and does not enter explicitly into the likelihood.  Instead, the
effective final state mistag is accounted for in the relative $B_s^0$ right-sign
and wrong-sign fraction.
For the $B_d$ contribution, the right-sign and wrong-sign decays are not 
treated separately and the overall event mistag rate factors in both
the inital state ($\eta_i$) and final state ($\eta_f^{B_d}$) mistags 
$\eta_{B_d}=\eta_{i} (1-\eta_{f}^{B_d}) + 
\eta_{f}^{B_d} (1-\eta_{i})$.  For $B^\pm$ and B Baryons, the overall event
mistag rate is obtained by subsituting $\eta_{f}^{B_d}$
in the previous equation 
with $\eta_{f}^{B^\pm}$ and $\eta_{f}^{BB}$.  The final state mistag rates are
listed in Table~\ref{tbl:etaf} and the various measured branching
ratios used in the calculation are listed in Table~\ref{tbl:phyparam}.
\begin{table}[hbt]
\begin{center}
\begin{tabular}{|c|c|c|} \hline
{}& $\eta_f$ (hadronic decay) & $\eta_f$ (semileptonic decays) \\ \hline \hline
$B_d$ &79.9$\%$ &88.5$\%$ \\
$B^{\pm}$ &77.4$\%$ &93.8$\%$ \\
B Baryons &92.1$\%$ &88.9$\%$ \\ \hline 
\end{tabular}
\caption {Final state mistag rates for the hadronic and the semileptonic 
(events with a lepton attached to the B vertex) decays.  For the
$B_d$, $B^\pm$, and B baryon events, the $D_s$ primarily comes
from the W decay, therefore, by definition, the final state mistag rates 
for those events are greater than 50$\%$.}
\label{tbl:etaf}
\end{center}
\end{table}

\subsubsection{Vertex Efficiency, Resolution Functions and Normalizations}
To complete the discussion on the likelihood function, two remaining
effects have to be addressed.  
First, the vertex efficiency
function is included in the physics functions to account for the lower
efficiency for reconstructing events with short proper time (events
close to the IP).  The efficiency functions are taken from the Monte Carlo
simulations and are parameterized by the function:
\begin{equation}
\epsilon (\tau) = P_1 \frac{1-e^{P_2\tau}}{1+e^{P_2\tau}}+P_3.
\label{eqn:eff1}
\end{equation}
Equation~(\ref{eqn:eff1}) models the vertex efficiency well for the
$B_s^0$, $B_d$ and B Baryons events.  However, for the $B^{\pm}$, 
the vertex efficiency is also expected to decline at large proper times 
due to the B vertex charge
requirement on the final sample(only candidates with Q=0,$\pm 1$ are kept).  
Therefore an additional exponential term ($P_4e^{-P_5\tau}$) is 
added to the function to model the behavior of the $B^\pm$ events. 
The coefficients ($P_{1-5}$) are listed in Table~\ref{tbl:vtxeff}
\begin{table}[hbt]
\begin{center}
\begin{tabular}{|c|c|c|c|c|c|} \hline
Category & $P_1$ &$P_2$ &$P_3$& $P_4$& $P_5$ \\ \hline \hline
$\epsilon_1,\epsilon_2$ & 0.148 & -6.856 & 0.033 &    &     \\
$\epsilon_3$            & 0.037 & -4.448 & 0.032 &    &     \\
$\epsilon_4$            & 0.061 & -0.985 & -1.209 & 1.243   & 0.007   \\
$\epsilon_5$            & 0.056 & -3.532 & 0.002 &    &     \\ \hline
\end{tabular}
\caption {Parameterizations for the vertex efficiency function.}
\label{tbl:vtxeff}
\end{center}
\end{table}

\begin{figure}[htb]
  \centering
  \epsfxsize8cm
  \leavevmode
  \epsfbox{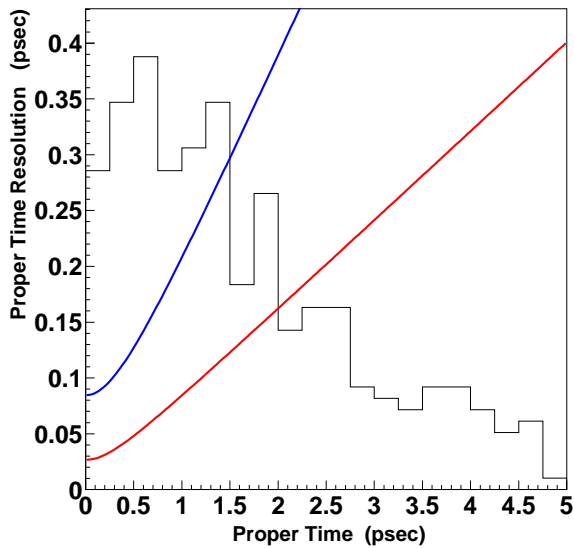}
  \baselineskip=12pt
  \caption{ \label{fig:sigtau}
  $\sigma_{\tau}$ distributions for $B_s^0 \rightarrow D_s^- X $ events.
  The lower red line is calculated using $\sigma_L(core)$ and 
  $\sigma_{\gamma\beta}(core)$.  The upper blue line is calculated using  
  $\sigma_L(tail)$ and $\sigma_{\gamma\beta}(tail)$.  The background
  histogram is the proper time distribution of the data events. }  
  \baselineskip=18pt
\end{figure}

The second effect concerns the proper time resolution.  The
effect is accounted for by introducing a resolution function in 
the convolution integrals, as shown in equations~(\ref{eqn:fbsrs}) to 
(\ref{eqn:fbb}).
The proper time resolution can be expressed in terms of the decay
length and boost resolutions ($\sigma_L$,$\sigma_{\gamma\beta}$) as in
\begin{equation}
\sigma_{\tau}(\tau,i,j) = \left[\left(\frac{\sigma_L^i}{\gamma\beta c}\right)^2
+ \left(\tau\,\frac{\sigma_{\gamma\beta}^j}{\gamma\beta}\right)^2\right]^{1/2}
\label{eqn:sigmat}\ ,
\end{equation}
where the indices (i and j) refer to core or tail.
The proper time resolution contains a constant term that depends on 
$\sigma_L$ and a term that rises linearly with proper time that depends 
on $\sigma_{\gamma\beta}$.  This
behavior is illustrated in Figure~\ref{fig:sigtau} for the $B_s^0$ events.
In practice, there are four $\sigma_{\tau}$ distributions given by
the various $\sigma_L$ and $\sigma_{\gamma\beta}$ core-tail combinations.
The resolution function is obtained by summing all four contributions
\begin{equation}
G(\tau,\tau_{rec})={\it f^2_{c}} G_{cc} + 
{\it f_{c}}{\it f_{t}}(G_{ct}+ G_{tc}) +
{\it f^2_{t}} G_{tt},
\end{equation}
with ${\it f_c}$=0.6, ${\it f_t}$=0.4, and $G_{i,j}$ defined as
\begin{equation}
G_{i,j}=\frac{1}{\sqrt{2\pi} \sigma_{\tau}(\tau,i,j)}exp 
\left[ {-(\tau-\tau_{rec})^2/
2\sigma_{\tau}^2(\tau,i,j)} \right].
\end{equation}

The complete likelihood function is normalized by dividing each physics 
term by a normalization constant.
The normalization constant ($N_{\it i}$) is 
calculated for category ${\it i}$ by integrating
the sum of the mixed and unmixed physics functions for source ${\it i}$ 
over all reconstructed proper times,
\begin{equation}
N_{\it i} = \int_{-\infty}^{+\infty} 
({F_{\it i}^{mixed}+F_{\it i}^{unmixed}}) \ d{\tau_{rec}}.
\end{equation}
The normalization is required to ensure that the weight of each
physics source in the likelihood function is not biased by the efficiency 
and the proper time resolution of the event.

\subsection{$D_s$+Tracks Amplitude Fit Results}
The physics parameters used in the amplitude fit are listed in
Table~\ref{tbl:phyparam}.  The systematic uncertainties are calculated
based on the formula~\cite{bmoser}
\begin{equation}
\sigma_A^{syst}=A^{new}-A^{nominal}+(1-A^{nominal})
\left( \frac{\sigma_A^{new}-\sigma_A^{nominal}}{\sigma_A^{nominal}} \right).
\end{equation}
\begin{table}[hbt]
\begin{center}
\begin{tabular}{|c|c|c|} \hline 
Parameter & Value and uncertainty  & Reference  \\  \hline \hline
$\tau_{B_s}$ &  $1.464\pm 0.057 \ ps$  & \cite{bbfrac1} \\
$\tau_{B_d}$ &  $1.562\pm 0.029 \ ps$  & \cite{bbfrac1} \\
$\tau_{B^{\pm}}$ &  $1.656\pm 0.025 \ ps$  & \cite{bbfrac1} \\
$\tau_{B Baryons}$ &  $1.208\pm 0.051 \ ps $  & \cite{bbfrac1} \\
$\Delta m_d$ & $0.476\pm 0.016 \ ps^{-1}$ & \cite{bbfrac1} \\ \hline
{\it f}($\overline{b} \rightarrow B_s^0$) &
 $0.1000\pm 0.012 $ &\cite{bbfrac1} \\
{\it f}($\overline{b} \rightarrow B_d^0, B^+$) & 
$0.4010\pm 0.010$&\cite{bbfrac1} \\
{\it f}($\overline{b} \rightarrow B \ Baryon$) & 
$0.0990\pm 0.017$&\cite{bbfrac1} 
\\ \hline
$R_b \cdot {\cal B}(b\rightarrow \overline{B_s^0}) \cdot 
{\cal B}(\overline{B_s^0}
 \rightarrow D_s^+X)\cdot {\cal B}(D_s^+\rightarrow \phi \pi^+)$&
$(6.21^{+0.73}_{-0.78})\times 10^{-4}$ & \cite{bbfrac1,bbfrac2} \\
${\cal B}(b\rightarrow W^- \rightarrow D_s^-)\cdot 
{\cal B}(D_s^- \rightarrow \phi \pi^-)$&
$(3.66\pm 0.45)\times 10^{-3}$ & \cite{bbfrac2} \\
${\cal B}(B_{d,u}\rightarrow D_s^{\pm} X)\cdot 
{\cal B}(D_s^- \rightarrow \phi \pi^-)$ &
$(3.71\pm 0.28)\times 10^{-3}$ & \cite{bbfrac1} \\
${\cal B}(B_{d,u}\rightarrow D_s^{-} X) / 
{\cal B}(B_{d,u}\rightarrow D_s^{\pm} X)$ &
$0.172\pm 0.083 $ & \cite{bbfrac1} \\ 
${\cal B}(\overline{c} \rightarrow D_s^-)\cdot 
{\cal B}(D_s^- \rightarrow \phi \pi^-)$&
$(3.4\pm 0.3)\times 10^{-3}$ & \cite{bbfrac1} \\ \hline
\end{tabular}
\caption {B lifetimes, $\Delta m_d$, B production fractions, and various 
branching ratios assumed in the amplitude fit. The uncertainties for the
branching ratios do not include uncertainty from 
Br($D_s \rightarrow \phi \pi$).  The right-sign $D_s$ production fraction 
($R_b \cdot {\cal B}(b\rightarrow \overline{B_s^0}) \cdot 
{\cal B}(\overline{B_s^0}
 \rightarrow D_s^+X)\cdot {\cal B}(D_s^+ \rightarrow \phi \pi^+)$) is
obtained by combining the direct measurement~\cite{bbfrac2} with
the semileptonic measurement~\cite{bbfrac1} assuming factorization.}
\label{tbl:phyparam}
\end{center}
\end{table}
The physics parameters are varied by $\pm 1 \sigma$ in the fit to
obtain the systematic uncertainties.
In addition, the uncertainity on the $D_s$ signal fraction (${\it f}_{D_s}$) 
is estimated from the $D_s$ mass fit and is varied by roughly 8$\%$.
The initial state tag probability is varied by $\pm 0.02$.
The decay length resolution and the boost resolution uncertainties have not
yet been studied in great detail and conservative estimates of 10$\%$ on
$\sigma_l$ and 30$\%$ on $\sigma_{\gamma\beta}$ are used.
A summary of the systematic uncertainties is given in Table~\ref{tbl:syst}  

\begin{figure}[p]
  \centering
  \epsfxsize10cm
  \leavevmode
  \epsfbox{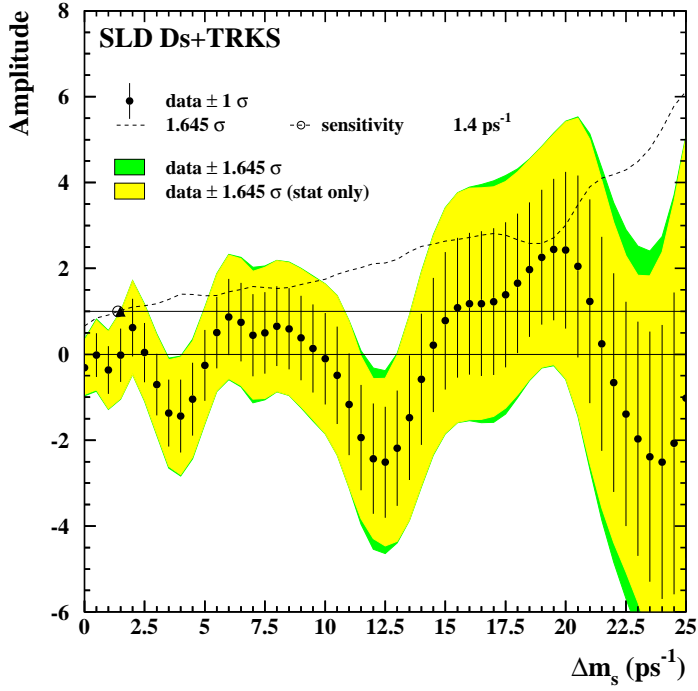}
  \caption{\it \label{fig:dsampfit}
  \baselineskip=12pt
  Ds+Tracks amplitude fit plot.}
  \baselineskip=18pt
\end{figure}

\begin{figure}[p]
  \centering
  \epsfxsize10cm
  \leavevmode
  \epsfbox{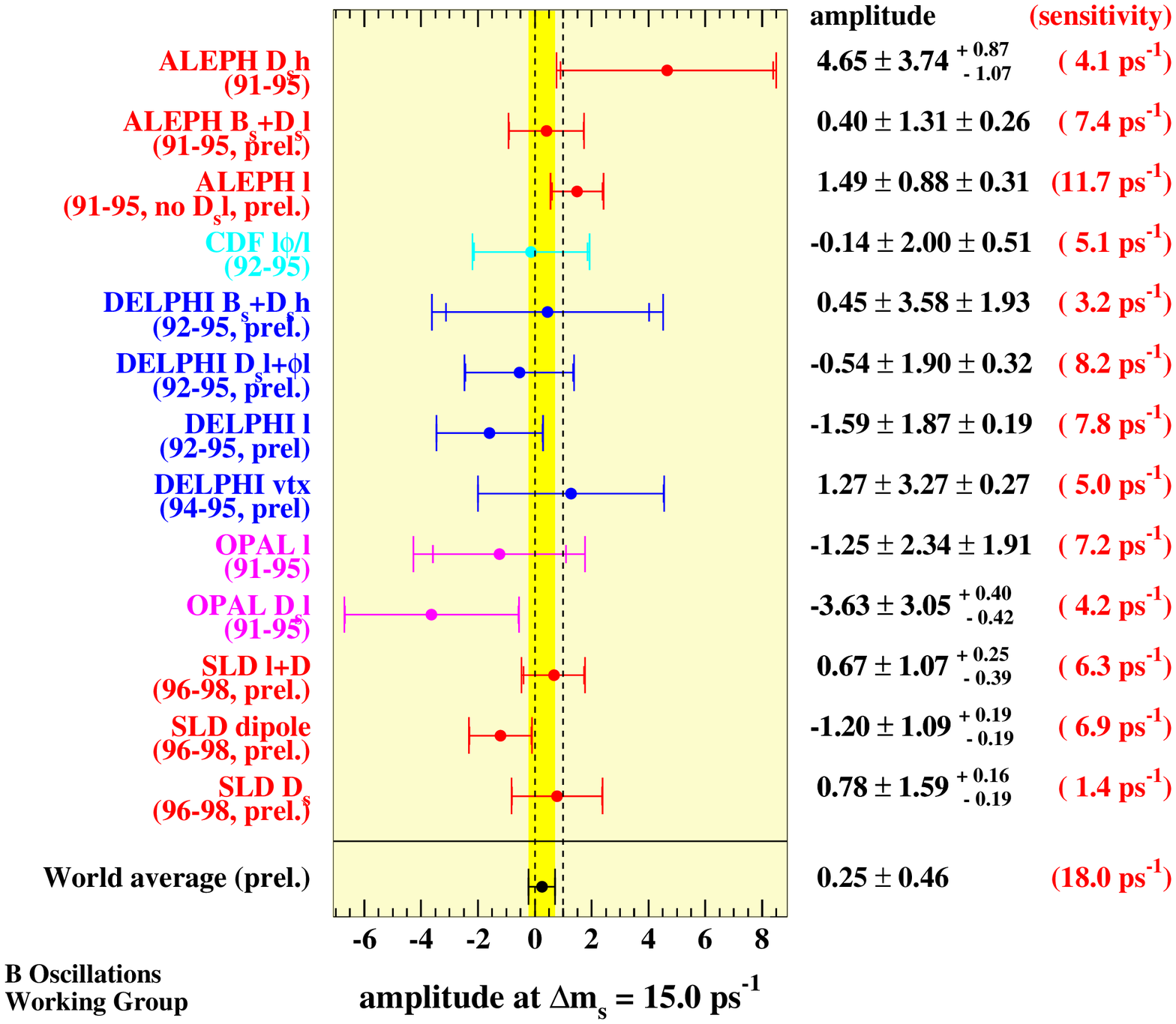}
  \caption{\it \label{fig:dms15}
  \baselineskip=12pt
  Amplitudes and errors at $\Delta m_s$=15$ps^{-1}$.}
  \baselineskip=18pt
\end{figure}

The resulting amplitude plot is shown in Figure~\ref{fig:dsampfit}.
The measured values of the amplitudes are consistent with zero within
the range of $\Delta m_s$ considered and no evidence of a signal is observed.
This analysis excludes the following values of the
$B_s^0 - \overline{B_s^0}$ mixing oscillation frequency:
$\Delta m_s < 1.5$ ps$^{-1}$,
$2.6 < \Delta m_s < 4.9$ ps$^{-1}$, and 
$10.8 < \Delta m_s < 13.5$ ps$^{-1}$ at the 95\% confidence level (C.L.).
The sensitivity at the 95\% C.L. is 1.4 $ps^{-1}$.
A comparison of the amplitudes and errors at $\Delta m_s$ = 15 $ps^{-1}$
for the various $B_s^0$ mixing analyses is shown in Figure~\ref{fig:dms15}  
(this analysis is listed as ``SLD Ds'' in the figure).

\newpage
\section{SLD Combination and Conclusion}
\begin{figure}[p]
  \centering
  \epsfxsize12cm
  \leavevmode
  \epsfbox{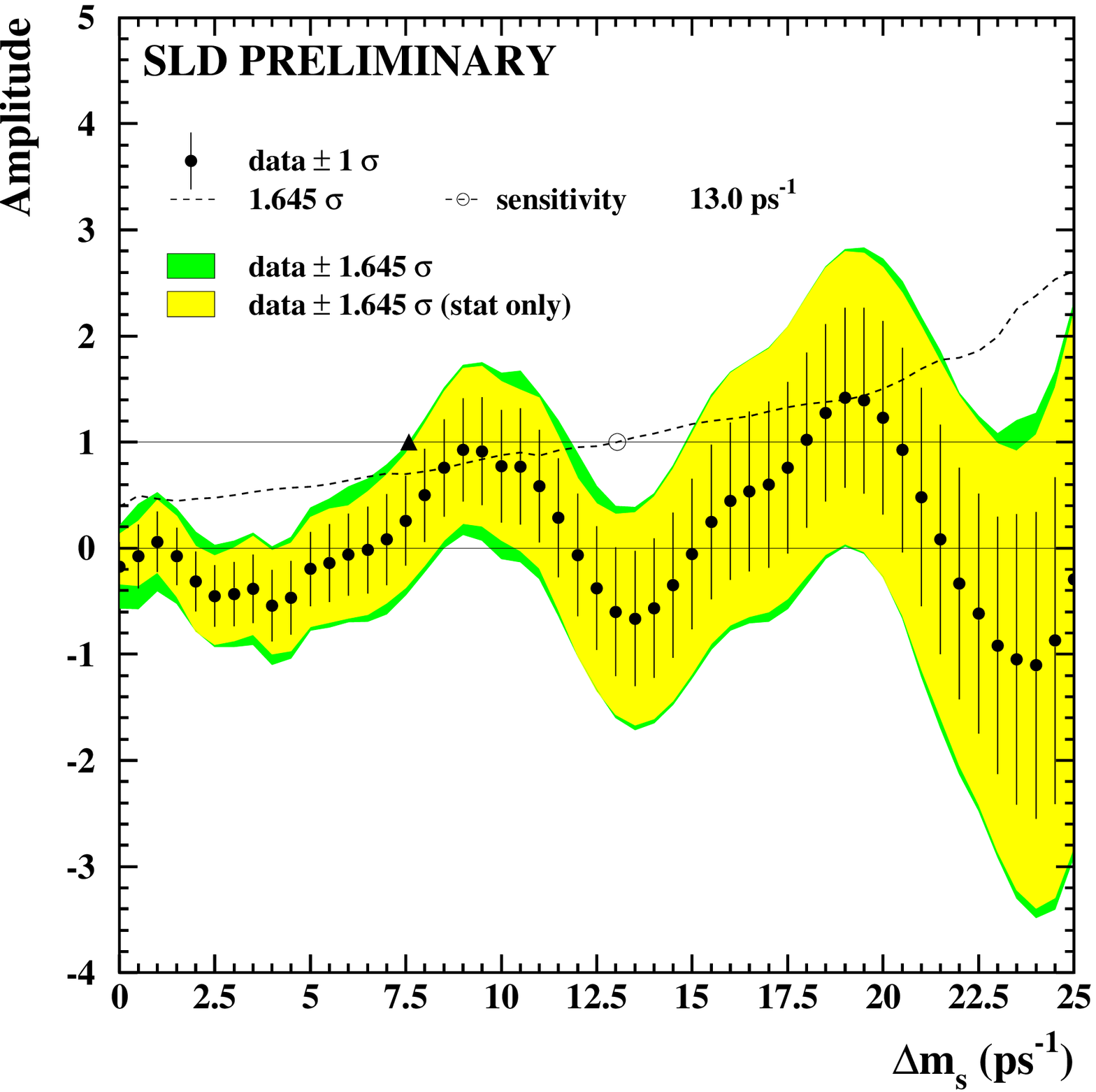}
  \caption{\it \label{fig:ampsld}
  Combined SLD amplitude plot.}
\end{figure}

The analysis presented in this paper is complementary to the two 
more inclusive analyses at the SLD: Charge Dipole and Lepton+D.
Charge Dipole and Lepton+D have a combined sensitivity of 
10 $ps^{-1}$~\cite{bdipo}.  
Figure~\ref{fig:ampsld} shows the amplitude fit for all three
analyses combined.  The combined plot takes into account the correlated
systematic uncertainties.  Furthermore, the samples were selected such
as to remove any statistical overlap between analyses.
No evidence of a signal is observed up to $\Delta m_s$ of 25 $ps^{-1}$ and
the excluded regions at the 95\% C.L. are: $\Delta m_s $ $<$ 7.6 $ps^{-1}$ and 
11.8 $<$ $\Delta m_s$ $<$ 14.8 $ps^{-1}$.  The SLD combined sensitivity at 
the 95\% C.L. is 13.0 $ps^{-1}$.
\begin{table}[p]
\begin{center}
\begin{tabular}{ccccc}
 \hline
 $\dms$                  & ~~5 ps$^{-1}$  & ~~10 ps$^{-1}$  & ~~15 ps$^{-1}$ 
                         & ~~20 ps$^{-1}$ \\
 \hline
 \vspace{0.1cm}
 Measured amplitude $A$  &       ~-0.260   &       ~-0.106   &     $0.781$ 
                         &      $-2.425$ \\
 \vspace{0.1cm}
 Statistical uncertainty ($\sigma_A^{stat}$)  &
    $\pm 0.813$  &   $\pm 1.058$  &   $\pm 1.591$
                         &   $\pm 1.817$  \\
 \vspace{0.1cm}
 Total systematic uncertainty ($\sigma_A^{syst})$       &
         $^{+0.163}_{-0.136}$ & $^{+0.129}_{-0.098}$ & $^{+0.161}_{-0.193}$
       & $^{+0.169}_{-0.221}$ \\
 \hline
 \vspace{0.2cm}
 $\tau_{B_s}$ &
         $^{+0.018}_{-0.019}$ & $^{+0.009}_{-0.009}$ & $^{+0.012}_{-0.013}$
       & $^{+0.006}_{-0.007}$ \\
 \vspace{0.1cm}
 $\tau_{B_d}$ &
         $^{-0.0006}_{+0.0006}$ & $^{-0.0005}_{+0.0005}$ & $^{-0.001}_{+0.001}$
       & $^{-0.0002}_{-0.0002}$ \\
 \vspace{0.1cm}
 $\tau_{B^\pm}$ &
         $^{-0.0007}_{+0.0007}$ & $^{-0.0002}_{+0.0002}$ & 
         $^{+0.00002}_{-0.00002}$ & $^{-0.001}_{+0.001}$ \\
 \vspace{0.1cm}
 $\tau_{B\ baryons}$ &
         $^{-0.00003}_{+0.00001}$ & $^{-0.00002}_{+0.00003}$ & 
         $^{+0.0001}_{-0.0001}$ & $^{+0.002}_{-0.002}$ \\
 \vspace{0.1cm}
 $\Delta m_d$ &
         $^{-0.0007}_{+0.0007}$ & $^{-0.0002}_{+0.0002}$ &
         $^{+0.00008}_{-0.00008}$ & $^{-0.00008}_{+0.00008}$ \\
 \vspace{0.1cm} 
  {\it f}($\overline{b} \rightarrow B_s^0$) &
         $^{+0.020}_{-0.019}$ & $^{+0.010}_{-0.010}$ &
         $^{+0.003}_{-0.003}$ & $^{-0.0007}_{+0.0003}$ \\
 \vspace{0.1cm}
  {\it f}($\overline{b} \rightarrow B\ Baryon$) &
         $^{+0.007}_{-0.007}$ & $^{+0.002}_{-0.002}$ &
         $^{+0.002}_{-0.002}$ & $^{-0.005}_{+0.005}$ \\
 \vspace{0.1cm}
  $\left( R_b \cdot {\cal B}(b\rightarrow \overline{B_s^0}) \cdot \right. $ &
         $^{-0.048}_{+0.069}$ & $^{-0.017}_{+0.025}$ &
         $^{-0.0006}_{-0.0014}$ & $^{-0.0068}_{+0.0007}$ \\
 $\left. {\cal B}(\overline{B_s^0}\rightarrow D_s^+X)\cdot 
 {\cal B}(D_s^+ \rightarrow \phi \pi^+)\right) $  & & & & \\
 \vspace{0.1cm}
 ${\cal B}(b\rightarrow W^- \rightarrow D_s^-)\cdot 
{\cal B}(D_s^- \rightarrow \phi \pi^-)$  &
         $^{+0.020}_{-0.020}$ & $^{+0.011}_{-0.010}$ &
         $^{+0.003}_{-0.003}$ & $^{-0.0008}_{+0.0003}$ \\
 \vspace{0.1cm}
 ${\cal B}(B_{d,u}\rightarrow D_s^{\pm} X)\cdot 
{\cal B}(D_s^- \rightarrow \phi \pi^-)$   &
         $^{+0.021}_{-0.021}$ & $^{+0.006}_{-0.006}$ &
         $^{-0.0009}_{+0.0008}$ & $^{+0.008}_{-0.009}$ \\
 \vspace{0.1cm}
  ${\cal B}(B_{d,u}\rightarrow D_s^{-} X) / 
{\cal B}(B_{d,u}\rightarrow D_s^{\pm} X)$  &
         $^{+0.014}_{-0.013}$ & $^{+0.024}_{-0.024}$ &
         $^{+0.020}_{-0.021}$ & $^{+0.033}_{-0.040}$ \\
 \vspace{0.1cm}
 ${\cal B}(\overline{c} \rightarrow D_s^-)\cdot 
{\cal B}(D_s^- \rightarrow \phi \pi^-)$  &
         $^{+0.006}_{-0.006}$ & $^{+0.004}_{-0.004}$ &
         $^{-0.0004}_{+0.0003}$ & $^{-0.005}_{+0.005}$ \\
 Decay length resolution      &
         $^{+0.004}_{-0.003}$ & $^{+0.039}_{-0.038}$ &
         $^{+0.015}_{-0.019}$ & $^{+0.027}_{-0.040}$ \\
 \vspace{0.1cm}
 Boost resolution      &
         $^{+0.052}_{-0.027}$ & $^{+0.075}_{-0.034}$ &
         $^{-0.166}_{+0.138}$ & $^{-0.147}_{+0.019}$ \\
 \vspace{0.1cm}
 ${\it f}_{D_s}$      &
         $^{+0.057}_{-0.055}$ & $^{+0.013}_{-0.011}$ &
         $^{+0.047}_{-0.051}$ & $^{+0.004}_{-0.006}$ \\
 \vspace{0.1cm}
 Initial state tag      &
         $^{-0.102}_{+0.118}$ & $^{-0.075}_{+0.089}$ &
         $^{-0.079}_{+0.061}$ & $^{-0.155}_{+0.162}$ \\
 \hline
\end{tabular}
\caption{Table of statistical and systematic uncertainties 
        for several $\dms$ values.}
\label{tbl:syst}
\end{center}
\end{table}

\section*{Acknowledgments}

        We thank the personnel of the SLAC accelerator department and
the technical staffs of our collaborating institutions for their outstanding
efforts.  This work was supported by the Department of Energy, the National
Science Foundataion, the Instituto Nazionale di Fisica of Italy, the
Japan-US Cooperative Research Project on High Energy Physics, and the
Science and Engineering Research Council of the United Kingdom.

\pagebreak

\newpage
%
%
%
\section*{$^{**}$ List of Authors}

\begin{center}
\def\iAOMORI{$^{(1)}$}
\def\iBRI{$^{(2)}$}
\def\iBRUN{$^{(3)}$}
\def\iBU{$^{(4)}$}
\def\iCOLO{$^{(5)}$}
\def\iCSU{$^{(6)}$}
\def\iFERR{$^{(7)}$}
\def\iFRAS{$^{(8)}$}
\def\iJHU{$^{(9)}$}
\def\iLBL{$^{(10)}$}
\def\iMASS{$^{(11)}$}
\def\iMISSI{$^{(12)}$}
\def\iMIT{$^{(13)}$}
\def\iMOSCOW{$^{(14)}$}
\def\iNAGO{$^{(15)}$}
\def\iOREG{$^{(16)}$}
\def\iOXF{$^{(17)}$}
\def\iPERU{$^{(18)}$}
\def\iRAL{$^{(19)}$}
\def\iRUTG{$^{(20)}$}
\def\iSLAC{$^{(21)}$}
\def\iSOONG{$^{(22)}$}
\def\iTENN{$^{(23)}$}
\def\iTOHO{$^{(24)}$}
\def\iUCSB{$^{(25)}$}
\def\iUCSC{$^{(26)}$}
\def\iVAND{$^{(27)}$}
\def\iWASH{$^{(28)}$}
\def\iWISC{$^{(29)}$}
\def\iYALE{$^{(30)}$}

  \baselineskip=.75\baselineskip  
\mbox{Kenji Abe\unskip,\iNAGO}
\mbox{Koya Abe\unskip,\iTOHO}
\mbox{T. Abe\unskip,\iSLAC}
\mbox{I. Adam\unskip,\iSLAC}
\mbox{H. Akimoto\unskip,\iSLAC}
\mbox{D. Aston\unskip,\iSLAC}
\mbox{K.G. Baird\unskip,\iMASS}
\mbox{C. Baltay\unskip,\iYALE}
\mbox{H.R. Band\unskip,\iWISC}
\mbox{T.L. Barklow\unskip,\iSLAC}
\mbox{J.M. Bauer\unskip,\iMISSI}
\mbox{G. Bellodi\unskip,\iOXF}
\mbox{R. Berger\unskip,\iSLAC}
\mbox{G. Blaylock\unskip,\iMASS}
\mbox{J.R. Bogart\unskip,\iSLAC}
\mbox{G.R. Bower\unskip,\iSLAC}
\mbox{J.E. Brau\unskip,\iOREG}
\mbox{M. Breidenbach\unskip,\iSLAC}
\mbox{W.M. Bugg\unskip,\iTENN}
\mbox{D. Burke\unskip,\iSLAC}
\mbox{T.H. Burnett\unskip,\iWASH}
\mbox{P.N. Burrows\unskip,\iOXF}
\mbox{A. Calcaterra\unskip,\iFRAS}
\mbox{R. Cassell\unskip,\iSLAC}
\mbox{A. Chou\unskip,\iSLAC}
\mbox{H.O. Cohn\unskip,\iTENN}
\mbox{J.A. Coller\unskip,\iBU}
\mbox{M.R. Convery\unskip,\iSLAC}
\mbox{V. Cook\unskip,\iWASH}
\mbox{R.F. Cowan\unskip,\iMIT}
\mbox{G. Crawford\unskip,\iSLAC}
\mbox{C.J.S. Damerell\unskip,\iRAL}
\mbox{M. Daoudi\unskip,\iSLAC}
\mbox{S. Dasu\unskip,\iWISC}
\mbox{N. de Groot\unskip,\iBRI}
\mbox{R. de Sangro\unskip,\iFRAS}
\mbox{D.N. Dong\unskip,\iSLAC}
\mbox{M. Doser\unskip,\iSLAC}
\mbox{R. Dubois\unskip,\iSLAC}
\mbox{I. Erofeeva\unskip,\iMOSCOW}
\mbox{V. Eschenburg\unskip,\iMISSI}
\mbox{S. Fahey\unskip,\iCOLO}
\mbox{D. Falciai\unskip,\iFRAS}
\mbox{J.P. Fernandez\unskip,\iUCSC}
\mbox{K. Flood\unskip,\iMASS}
\mbox{R. Frey\unskip,\iOREG}
\mbox{E.L. Hart\unskip,\iTENN}
\mbox{K. Hasuko\unskip,\iTOHO}
\mbox{S.S. Hertzbach\unskip,\iMASS}
\mbox{M.E. Huffer\unskip,\iSLAC}
\mbox{X. Huynh\unskip,\iSLAC}
\mbox{M. Iwasaki\unskip,\iOREG}
\mbox{D.J. Jackson\unskip,\iRAL}
\mbox{P. Jacques\unskip,\iRUTG}
\mbox{J.A. Jaros\unskip,\iSLAC}
\mbox{Z.Y. Jiang\unskip,\iSLAC}
\mbox{A.S. Johnson\unskip,\iSLAC}
\mbox{J.R. Johnson\unskip,\iWISC}
\mbox{R. Kajikawa\unskip,\iNAGO}
\mbox{M. Kalelkar\unskip,\iRUTG}
\mbox{H.J. Kang\unskip,\iRUTG}
\mbox{R.R. Kofler\unskip,\iMASS}
\mbox{R.S. Kroeger\unskip,\iMISSI}
\mbox{M. Langston\unskip,\iOREG}
\mbox{D.W.G. Leith\unskip,\iSLAC}
\mbox{V. Lia\unskip,\iMIT}
\mbox{C. Lin\unskip,\iMASS}
\mbox{G. Mancinelli\unskip,\iRUTG}
\mbox{S. Manly\unskip,\iYALE}
\mbox{G. Mantovani\unskip,\iPERU}
\mbox{T.W. Markiewicz\unskip,\iSLAC}
\mbox{T. Maruyama\unskip,\iSLAC}
\mbox{A.K. McKemey\unskip,\iBRUN}
\mbox{R. Messner\unskip,\iSLAC}
\mbox{K.C. Moffeit\unskip,\iSLAC}
\mbox{T.B. Moore\unskip,\iYALE}
\mbox{M. Morii\unskip,\iSLAC}
\mbox{D. Muller\unskip,\iSLAC}
\mbox{V. Murzin\unskip,\iMOSCOW}
\mbox{S. Narita\unskip,\iTOHO}
\mbox{U. Nauenberg\unskip,\iCOLO}
\mbox{H. Neal\unskip,\iYALE}
\mbox{G. Nesom\unskip,\iOXF}
\mbox{N. Oishi\unskip,\iNAGO}
\mbox{D. Onoprienko\unskip,\iTENN}
\mbox{L.S. Osborne\unskip,\iMIT}
\mbox{R.S. Panvini\unskip,\iVAND}
\mbox{C.H. Park\unskip,\iSOONG}
\mbox{I. Peruzzi\unskip,\iFRAS}
\mbox{M. Piccolo\unskip,\iFRAS}
\mbox{L. Piemontese\unskip,\iFERR}
\mbox{R.J. Plano\unskip,\iRUTG}
\mbox{R. Prepost\unskip,\iWISC}
\mbox{C.Y. Prescott\unskip,\iSLAC}
\mbox{B.N. Ratcliff\unskip,\iSLAC}
\mbox{J. Reidy\unskip,\iMISSI}
\mbox{P.L. Reinertsen\unskip,\iUCSC}
\mbox{L.S. Rochester\unskip,\iSLAC}
\mbox{P.C. Rowson\unskip,\iSLAC}
\mbox{J.J. Russell\unskip,\iSLAC}
\mbox{O.H. Saxton\unskip,\iSLAC}
\mbox{T. Schalk\unskip,\iUCSC}
\mbox{B.A. Schumm\unskip,\iUCSC}
\mbox{J. Schwiening\unskip,\iSLAC}
\mbox{V.V. Serbo\unskip,\iSLAC}
\mbox{G. Shapiro\unskip,\iLBL}
\mbox{N.B. Sinev\unskip,\iOREG}
\mbox{J.A. Snyder\unskip,\iYALE}
\mbox{H. Staengle\unskip,\iCSU}
\mbox{A. Stahl\unskip,\iSLAC}
\mbox{P. Stamer\unskip,\iRUTG}
\mbox{H. Steiner\unskip,\iLBL}
\mbox{D. Su\unskip,\iSLAC}
\mbox{F. Suekane\unskip,\iTOHO}
\mbox{A. Sugiyama\unskip,\iNAGO}
\mbox{S. Suzuki\unskip,\iNAGO}
\mbox{M. Swartz\unskip,\iJHU}
\mbox{F.E. Taylor\unskip,\iMIT}
\mbox{J. Thom\unskip,\iSLAC}
\mbox{E. Torrence\unskip,\iMIT}
\mbox{T. Usher\unskip,\iSLAC}
\mbox{J. Va'vra\unskip,\iSLAC}
\mbox{R. Verdier\unskip,\iMIT}
\mbox{D.L. Wagner\unskip,\iCOLO}
\mbox{A.P. Waite\unskip,\iSLAC}
\mbox{S. Walston\unskip,\iOREG}
\mbox{A.W. Weidemann\unskip,\iTENN}
\mbox{E.R. Weiss\unskip,\iWASH}
\mbox{J.S. Whitaker\unskip,\iBU}
\mbox{S.H. Williams\unskip,\iSLAC}
\mbox{S. Willocq\unskip,\iMASS}
\mbox{R.J. Wilson\unskip,\iCSU}
\mbox{W.J. Wisniewski\unskip,\iSLAC}
\mbox{J.L. Wittlin\unskip,\iMASS}
\mbox{M. Woods\unskip,\iSLAC}
\mbox{T.R. Wright\unskip,\iWISC}
\mbox{R.K. Yamamoto\unskip,\iMIT}
\mbox{J. Yashima\unskip,\iTOHO}
\mbox{S.J. Yellin\unskip,\iUCSB}
\mbox{C.C. Young\unskip,\iSLAC}
\mbox{H. Yuta\unskip.\iAOMORI}

\it
  \vskip \baselineskip                   
  \centerline{(The SLD Collaboration)}   
  \vskip \baselineskip        
  \baselineskip=.75\baselineskip   
\iAOMORI
  Aomori University, Aomori , 030 Japan, \break
\iBRI
  University of Bristol, Bristol, United Kingdom, \break
\iBRUN
  Brunel University, Uxbridge, Middlesex, UB8 3PH United Kingdom, \break
\iBU
  Boston University, Boston, Massachusetts 02215, \break
\iCOLO
  University of Colorado, Boulder, Colorado 80309, \break
\iCSU
  Colorado State University, Ft. Collins, Colorado 80523, \break
\iFERR
  INFN Sezione di Ferrara and Universita di Ferrara, I-44100 Ferrara, Italy, \break
\iFRAS
  INFN Lab. Nazionali di Frascati, I-00044 Frascati, Italy, \break
\iJHU
  Johns Hopkins University,  Baltimore, Maryland 21218-2686, \break
\iLBL
  Lawrence Berkeley Laboratory, University of California, Berkeley, California 94720, \break
\iMASS
  University of Massachusetts, Amherst, Massachusetts 01003, \break
\iMISSI
  University of Mississippi, University, Mississippi 38677, \break
\iMIT
  Massachusetts Institute of Technology, Cambridge, Massachusetts 02139, \break
\iMOSCOW
  Institute of Nuclear Physics, Moscow State University, 119899, Moscow Russia, \break
\iNAGO
  Nagoya University, Chikusa-ku, Nagoya, 464 Japan, \break
\iOREG
  University of Oregon, Eugene, Oregon 97403, \break
\iOXF
  Oxford University, Oxford, OX1 3RH, United Kingdom, \break
\iPERU
  INFN Sezione di Perugia and Universita di Perugia, I-06100 Perugia, Italy, \break
\iRAL
  Rutherford Appleton Laboratory, Chilton, Didcot, Oxon OX11 0QX United Kingdom, \break
\iRUTG
  Rutgers University, Piscataway, New Jersey 08855, \break
\iSLAC
  Stanford Linear Accelerator Center, Stanford University, Stanford, California 94309, \break
\iSOONG
  Soongsil University, Seoul, Korea 156-743, \break
\iTENN
  University of Tennessee, Knoxville, Tennessee 37996, \break
\iTOHO
  Tohoku University, Sendai 980, Japan, \break
\iUCSB
  University of California at Santa Barbara, Santa Barbara, California 93106, \break
\iUCSC
  University of California at Santa Cruz, Santa Cruz, California 95064, \break
\iVAND
  Vanderbilt University, Nashville,Tennessee 37235, \break
\iWASH
  University of Washington, Seattle, Washington 98105, \break
\iWISC
  University of Wisconsin, Madison,Wisconsin 53706, \break
\iYALE
  Yale University, New Haven, Connecticut 06511. \break

\rm
%

\end{center}

\enddocument